\documentclass[pra,superscriptaddress,showpacs,nobalancelastpage]{revtex4}

\usepackage{graphicx}
\begin{document}

\title{Continuous variable entanglement and quantum state teleportation
between optical and macroscopic vibrational modes through radiation pressure}

\author{
Stefano Pirandola, Stefano Mancini,
David Vitali,
and Paolo Tombesi}
\affiliation{
INFM, Dipartimento di Fisica,
Universit\`a di Camerino,
via Madonna delle Carceri, I-62032 Camerino, Italy}

\date{\today}

\begin{abstract}
We study an isolated, perfectly reflecting, mirror illuminated by an intense laser pulse. We show that
the resulting radiation pressure efficiently entangles a mirror vibrational mode with the two reflected optical sideband modes
of the incident carrier beam. The entanglement of the resulting three-mode state is studied in detail and it is shown to be
robust against the mirror mode temperature. We then show how this continuous variable entanglement can be profitably used to
teleport an unknown quantum state of an optical mode
onto the vibrational mode of the mirror.
\end{abstract}

\pacs{Pacs No: 03.67.-a, 42.50.Vk, 03.65.Ud}

\maketitle

\section{Introduction}

Entanglement is the property possessed by a multipartite quantum systems when it is in a state that cannot be factorized into a product of states
or a mixture of these products. In an entangled state the various parties share nonclassical and nonlocal correlations,
which are at the basis of the attractive and counterintuitive aspects of quantum mechanics.
The study of the properties of entanglement has been recently characterized by an impressive development caused by the discovery that
it represents a valuable resource, allowing tasks which are impossible if only separable (unentangled) states are used. In particular,
the entanglement shared by two parties significantly increases their communication capabilities.
Perhaps the most famous example is quantum state teleportation,
i.e. the possibility to send the information contained in a quantum state in a reliable way \cite{BEN93}.
Other important examples are quantum dense coding, i.e., the possibility to double the classical information transmitted along a channel \cite{WIE93},
and the possibility to realize quantum key distribution for cryptography \cite{EKE91}. All these schemes for
quantum communication have been originally developed for qubits. However, these schemes have been extended quite soon to continuous variable (CV) systems,
i.e., quantum systems characterized by observables with a fully continuum spectrum, like position and momentum of a particle,
or the quadratures of an e.m. field. In fact, it has been shown, first theoretically \cite{VAI94,BRA98} and then experimentally \cite{FUR98,BOW03}
how to teleport the quantum state of an e.m. field. Later, most of the quantum communication protocols have been extended to the
CV case: dense coding \cite{DCOD}, quantum key distribution \cite{CVCRYP} (experimentally demonstrated recently in \cite{GRA03}),
and also the possibility to perform
CV quantum computation has been discussed \cite{CVQCOMP}. The interest in CV quantum communication protocols
has considerably increased in the last years due to the relative simplicity with which it is possible to generate and manipulate
the CV entangled states of the electromagnetic field. The prototype entangled state for CV systems is the
simultaneous eigenstate of total momentum and relative distance of two particles, used by Einstein, Podolsky and Rosen (EPR) in their well-known
argument \cite{EIN35}. The recent development of
CV quantum information is mainly due to the fact that its finite-energy version for a two-optical mode system
in which the phase and amplitude field quadrature play the role of position and momentum of the particles is just the {\em two-mode squeezed} (TMS) state,
which can be easily generated by a non degenerate parametric amplifier \cite{QO94}. This means that CV entanglement can be
generated in a relatively easy way; also its manipulation is not particularly difficult because it requires only linear passive
elements as beam splitters and phase shifters. Finally the detection of these entangled state is easy because the Bell measurements
(measurements in an entangled basis) in the CV case reduce to homodyne measurements of appropriate quadratures.

However, in any future CV quantum network, it is necessary to have not only light
fastly connecting the various nodes, but also physical systems able to store in a stable way and locally manipulate quantum information.
This means considering media able to interact and exchange quantum information with optical fields in an efficient way. Moreover
this medium should possess entangled states particularly robust and stable against the effects of external noise and imperfections,
so that they can be locally manipulated for all the time needed. In the recent literature, various candidates have been
proposed for storing CV quantum information. A relatively advanced scheme is the one based on collective atomic spin states
\cite{POL,JUL01}, where the involved variable is the total spin, which is actually discrete, but can be safely treated as continuous in the limit of a
large number of atoms. In such a case, the presence of entanglement has been already experimentally demonstrated \cite{JUL01}.
Other schemes are based on mechanical degrees of freedom rather than atomic internal states as, for example,
vibrational modes of trapped and cooled ions or atoms \cite{PARK}, or localized acoustic waves in mirrors \cite{PRL02,PRL03}.
The proposal based on trapped atoms or ions requires the use of traps in conjunction with high-Q cavities, which is technically
difficult and for which only preliminary steps have been demonstrated recently \cite{CAV}. Here we shall discuss in detail
the scheme in which CV quantum information is encoded in appropriate acoustic modes of a mirror
and for which it has been recently demonstrated the possibility to realize quantum teleportation \cite{PRL03}.
The radiation pressure of an optical beam incident on a mirror realizes an effective coupling between the electromagnetic
modes reflected by the mirror and the vibrational mode of the latter. This coupling can be tuned by varying the intensity of the
incident beam. Even though of classical origin, radiation pressure can have important quantum effects: for example, it has been shown
that it may be used to generate nonclassical states of both the radiation field \cite{FAB,PRA94},
and the motional degree of freedom of the mirror \cite{PRA97,marshall}.

Moreover, the optomechanical coupling with a vibrating mirror
yields an effective nonlinearity for the radiation modes impinging on it. For example
optical bistability has been demonstrated in a cavity with a movable mirror \cite{bistab}.
This implies that the mirror, reacting to the radiation pressure, realizes an effective
coupling between different optical modes impinging on it. The dynamics of these interacting modes in the case of
a driven multi-mode cavity with a movable and perfectly reflecting mirror has been
analyzed in detail in Refs.~\cite{GMT,ALE,SILVIA}, which studied in particular how CV
entanglement between the modes can be generated and characterized. Here, differently from
Refs.~\cite{GMT,ALE,SILVIA}, we do not consider an optical cavity but only a single mirror, illuminated by an intense and highly monochromatic optical
beam. Moreover, we shall not consider only the entanglement between different optical modes but also between the reflected optical modes
and the mirror vibrational modes. In particular we shall see that acousto-optical entanglement can be used
to teleport the state of an optical mode onto a mirror acoustic mode. This fact opens up the possibility to use a
vibrational mode of a mirror as a CV quantum memory and also shows that quantum mechanics
can have unexpected manifestations at macroscopic level, involving collective vibrational modes with an effective
mass of the order of $\mu$g \cite{PRL03,marshall}.

The paper is organized as follows. In Section II we introduce the system under study, i.e. an isolated mirror illuminated by an intense
an quasi-monochromatic light pulse.
Under suitable approximations usually satisfied in realistic experimental situations, we derive an effective Hamiltonian
for three interacting bosonic modes, the two optical sidebands of the driving mode and one vibrational mode of the mirror.
In Section III we exactly solve the three-mode dynamics
and in Section IV we perform an extensive study of the CV entanglement between these modes. We shall see that
under particular conditions, radiation pressure becomes a source of two-mode squeezing between the two sideband modes. In Section V we then
show how the optomechanical entanglement between light and an acoustic mode can be used to teleport a CV state of an optical mode
onto the mirror vibrational mode. We shall consider and compare two slightly different protocols, which are both a three-mode adaptation
of the protocol by Vaidman, \cite{VAI94} and by Braunstein and Kimble \cite{BRA98}. Section VI is for concluding remarks.

\section{The system}

The possibility to generate CV entanglement through radiation pressure has been already
shown adopting schemes involving a cavity. In Refs.~\cite{GMT,ALE,SILVIA} it has been shown how radiation pressure entangles different modes
of a driven multi-mode cavity, while Ref.~\cite{PRL02} has shown how the radiation pressure of an intra-cavity mode entangles the vibrational modes
of two oscillating mirrors of a ring cavity. Here we shall consider a simpler scheme with only one vibrating mirror,
illuminated by the continuum of travelling electromagnetic modes in free space. This very simple scheme
may be useful for quantum communication purposes and moreover it will allow us to study the fundamental
aspects of the optomechanical entanglement generated by radiation pressure.

We study for simplicity a perfectly reflecting mirror
and consider only the motion and elastic deformations
taking place along the spatial direction $x$,
orthogonal to its reflecting surface.
In the limit of small mirror displacements, and in the interaction
picture with respect to the free Hamiltonian of the electromagnetic field
and of the mirror displacement field ${\hat x({\bf r},t)}$
(${\bf r}$ is the two-dimensional coordinate on the mirror surface), one has the
following Hamiltonian
 \cite{PIN99}
\begin{eqnarray}
    {\hat H}&=&-\int\,d^{2}{\bf r}\,
    {\hat P}({\bf r},t){\hat x}({\bf r},t)\,,
    \label{eq:Hini}
\end{eqnarray}
where ${\hat P({\bf r},t)}$ is the radiation pressure force \cite{SAM95}.
All the continuum of electromagnetic modes
with positive longitudinal wave vector $q$, transverse
wave vector ${\bf k}$, and frequency $\omega=\sqrt{c^{2}(k^{2}+q^{2})}$
($c$ being the light speed in the vacuum)
contributes to the radiation pressure force.
Following Ref.~\cite{SAM95}, and considering linearly polarized radiation
with the electric field parallel to the mirror surface, we have
\begin{eqnarray}
        {\hat P}({\bf r},t)&=&-\frac{\hbar}{8\pi^{3}}
        \int d{\bf k}\int dq\int d{\bf k'}\int dq'
        \frac{c^{2}qq'}{\sqrt{\omega\omega'}}
        ({\bf u}_{k}\cdot{\bf u}_{k'}){\bf u}_{q}
        \nonumber\\
        &\times&
        \left\{{\hat a}({\bf k},q){\hat a}({\bf k}',q')
        \exp[-i(\omega+\omega')t
        +i({\bf k}+{\bf k}')\cdot{\bf r}]\right.
        \nonumber\\
        &&\left.
        +{\hat a}^{\dag}({\bf k},q){\hat a}^{\dag}({\bf k}',q')
        \exp[i(\omega+\omega')t
        -i({\bf k}+{\bf k}')\cdot{\bf r}]\right.
        \nonumber\\
        &&\left.
        +{\hat a}({\bf k},q){\hat a}^{\dag}({\bf k}',q')
        \exp[-i(\omega-\omega')t
        +i({\bf k}-{\bf k}')\cdot{\bf r}]\right.
        \nonumber\\
        &&\left.
        +{\hat a}^{\dag}({\bf k},q){\hat a}({\bf k}',q')
        \exp[i(\omega-\omega')t
        -i({\bf k}-{\bf k}')\cdot{\bf r}]
        \right\}\,,
        \label{eq:P}
\end{eqnarray}
where ${\hat a}({\bf k},q)$ are the continuous mode destruction
operators, obeying the commutation
relations
\begin{equation}
\left[{\hat a}({\bf k},q),{\hat a}({\bf k}',q')\right]
=\delta({\bf k}-{\bf k}')\delta(q-q')\,,
\end{equation}
and ${\bf u}_{k}$, ${\bf u}_{q}$ denote dimensionless
unit vectors parallel to ${\bf k}$, $q$ respectively.

The mirror displacement ${\hat x({\bf r},t)}$ is generally given by a
superposition of many acoustic modes \cite{PIN99};
however, a single vibrational mode description can be adopted whenever
detection is limited to a frequency bandwidth
including a single mechanical resonance.
In particular, focused light beams are able to excite
Gaussian acoustic modes, in which only a small portion of the mirror,
localized at its center, vibrates. These modes have a small
waist $w$, a large mechanical quality
factor $Q$, a small effective mass $M$ \cite{PIN99}, and
the simplest choice is to choose the fundamental Gaussian mode with
frequency $\Omega$ and annihilation operator $b$, i.e.,
\begin{equation}
    {\hat x}({\bf r},t)=\sqrt{\frac{\hbar}{2M\Omega}}
    \left[{\hat b} e^{-i\Omega t}+{\hat b}^{\dag}e^{i\Omega t}\right]
    \exp(-r^{2}/w^{2})\,.
    \label{eq:x}
\end{equation}
By inserting Eqs.~(\ref{eq:P}) and (\ref{eq:x}) into Eq.~(\ref{eq:Hini})
and integrating over the mirror surface, one gets
\begin{eqnarray}
        {\hat H}&=&-\frac{\hbar
        w^{2}}{8\pi^{2}}\sqrt{\frac{\hbar}{2M\Omega}}
        \int d{\bf k}\int dq\int d{\bf k'}\int dq'
        \frac{c^{2}qq'}{\sqrt{\omega\omega'}}
        ({\bf u}_{k}\cdot{\bf u}_{k'}){\bf u}_{q}
        \nonumber\\
        &\times&
        \left\{{\hat a}({\bf k},q){\hat a}({\bf k}',q')
        \exp[-i(\omega+\omega')t
        -({\bf k}+{\bf k}')^{2}w^{2}/4]\right.
        \nonumber\\
        &&\left.
        +{\hat a}^{\dag}({\bf k},q){\hat a}^{\dag}({\bf k}',q')
        \exp[i(\omega+\omega')t
        -({\bf k}+{\bf k}')^{2}w^{2}/4]\right.
        \nonumber\\
        &&\left.
        +{\hat a}({\bf k},q){\hat a}^{\dag}({\bf k}',q')
        \exp[-i(\omega-\omega')t
        -({\bf k}-{\bf k}')^{2}w^{2}/4]\right.
        \nonumber\\
        &&\left.
        +{\hat a}^{\dag}({\bf k},q){\hat a}({\bf k}',q')
        \exp[i(\omega-\omega')t
        -({\bf k}-{\bf k}')^{2}w^{2}/4]
        \right\}\times\left\{{\hat b}e^{-i\Omega t}
        +{\hat b}^{\dag}e^{i\Omega t}\right\}\,.
        \label{eq:Hint1}
\end{eqnarray}
The acoustical waist $w$ is typically much larger than
optical wavelengths \cite{PIN99}, and therefore we can approximate
$\exp\left\{-({\bf k}\pm {\bf k'})^{2}w^{2}/4\right\}w^{2}/4\pi \simeq
\delta({\bf k}\pm {\bf k'})$ and then integrate Eq.~(\ref{eq:Hint1})
over ${\bf k'}$, obtaining
\begin{eqnarray}
        {\hat H}&=&-\frac{\hbar}{2\pi}\sqrt{\frac{\hbar}{2M\Omega}}
        \int d{\bf k}\int dq\int dq'
        \frac{c^{2}qq'}{\sqrt{\omega\omega'}}
        \nonumber\\
        &\times&
        \left\{-{\hat a}({\bf k},q){\hat a}(-{\bf
        k},q')\exp[-i(\omega+\omega')t]
        \right.
        \nonumber\\
        &&\left.
        -{\hat a}^{\dag}({\bf k},q){\hat a}^{\dag}(-{\bf
        k},q')\exp[i(\omega+\omega')t]\right.
        \nonumber\\
        &&\left.
        +{\hat a}({\bf k},q){\hat a}^{\dag}({\bf
        k},q')\exp[-i(\omega-\omega')t]\right.
        \nonumber\\
        &&\left.
        +{\hat a}^{\dag}({\bf k},q){\hat a}({\bf k},q')
        \exp[i(\omega-\omega')t]
        \right\}\times\left\{{\hat b}e^{-i\Omega t}+{\hat
        b}^{\dag}e^{i\Omega t}\right\}\,.
        \label{eq:Hint2}
\end{eqnarray}
This is equivalent to consider a very large mirror surface which implies that the transverse momentum of photons is conserved upon reflection.

Here we are considering the time evolution of the continuum of optical modes impinging on the mirror and travelling at the speed of light.
The experimental time resolution $\tau$ with which this dynamics is observed is given by the inverse of the detection bandwidth $\Delta\nu_{det}$.
Therefore we have to average the Hamiltonian of Eq.~(\ref{eq:Hint2}) over this time resolution $\tau$.
We assume $ \Delta\nu_{det} \ll \Omega $ which means
neglecting all the terms oscillating in time faster than the mechanical
frequency $\Omega$, yielding the following replacements in
Eq.~(\ref{eq:Hint2})
\begin{equation}
\exp\left\{\pm i(\omega' \pm \omega \pm \Omega)t\right\} \rightarrow
\frac{2\pi}{\tau}\delta(\omega' \pm \omega \pm \Omega).
\end{equation}
Since $\omega$ and $\omega'$ are positive and $\Omega$ is much
smaller than typical optical frequencies, the two terms
$\delta(\omega' + \omega \pm \Omega)$ give no contribution, while the
other two terms can be rewritten as
\begin{equation}
\frac{2\pi}{\tau}\delta(\omega' - \omega \pm \Omega) =
2\pi\Delta\nu_{det}\delta(q' -
\bar{q}_{\pm})\frac{\omega'(\bar{q}_{\pm})}{c^{2}\bar{q}_{\pm}},
\end{equation}
where $\bar{q}_{\pm}=\sqrt{(\omega \pm \Omega)^{2}/c^{2}-k^{2}}$.
Integrating over $q'$ we get
\begin{eqnarray}
        {\hat H}&=&-\hbar\Delta\nu_{det}\sqrt{\frac{\hbar}{2M\Omega}}
        \int d{\bf k}\int dq
        \frac{q}{\sqrt{\omega}}
        \left\{{\hat a}({\bf k},q){\hat a}^{\dag}\left({\bf
        k},\bar{q}_{+}\right){\hat b}\;
        \sqrt{\omega+\Omega}
        +{\hat a}({\bf k},q){\hat a}^{\dag}\left({\bf k},
        \bar{q}_{-}\right){\hat b}^{\dag}\sqrt{\omega-\Omega}
        \right.
        \nonumber\\
        &&\left.
        +{\hat a}^{\dag}({\bf k},q){\hat a}\left({\bf k},
        \bar{q}_{+}\right){\hat b}^{\dag}
        \sqrt{\omega+\Omega}
        +{\hat a}^{\dag}({\bf k},q){\hat a}\left({\bf k},
        \bar{q}_{-}\right){\hat b}\sqrt{\omega-\Omega}
        \right\}\,,
        \label{eq:Hint3}
\end{eqnarray}
where we have used the fact that $\omega'(\bar{q}_{\pm})=\omega \pm
\Omega$.

We now consider the situation where the radiation field incident on
the mirror is characterized by an intense, quasi-monochromatic,
laser field with transversal
wave vector ${\bf k_{0}}$, longitudinal wave vector $q_{0}$,
cross-sectional area $A$, and power ${\wp}$. Since this component is
very intense, it can be
treated as classical and one can approximate
the bosonic operators ${\hat a}({\bf k},q)$ describing the incident field in Eq.~(\ref{eq:Hint3}), with the c-number field
\begin{equation}
\alpha({\bf k},q) = -i\sqrt{\frac{(2\pi)^{3}{\wp}}{\hbar \omega_{0}cA}}
\delta({\bf k}-{\bf k_{0}})\delta(q-q_{0})\,,
\label{intenso}
\end{equation}
describing the classical coherent amplitude of the incident field, (it is $\omega_{0}=c\sqrt{{\bf k_{0}}^{2}+q_{0}^{2}}$ in Eq.~(\ref{intenso})), while
keeping a quantum description for the reflected field ${\hat a}({\bf k},\bar{q}_{\pm})$.
Inserting Eq.~(\ref{intenso}) into Eq.~(\ref{eq:Hint3}), the only nonvanishing terms are those
involving two back-scattered waves, that is, the sidebands of the driving
beam at frequencies
$\omega_{0}\pm \Omega$, as described by
\begin{eqnarray}
        {\hat H}&=& i\hbar\Delta\nu_{det}\sqrt{\frac{\hbar}{2M\Omega}}
        q_{0} \sqrt{\frac{{\wp}}{\hbar \omega_{0}cA}}
        \left\{\sqrt{\frac{\omega_{0}+\Omega}{\omega_{0}}}
        {\hat a}^{\dag}\left({\bf
        k_{0}},\bar{q}_{+}\right){\hat b}
        +\sqrt{\frac{\omega_{0}-\Omega}{\omega_{0}}}
        {\hat a}^{\dag}\left({\bf
        k_{0}},\bar{q}_{-}\right){\hat b}^{\dagger}
        \right.
        \nonumber\\
        &&\left.
        -\sqrt{\frac{\omega_{0}+\Omega}{\omega_{0}}}
        {\hat a}\left({\bf
        k_{0}},\bar{q}_{+}\right){\hat b}^{\dagger}
        -\sqrt{\frac{\omega_{0}-\Omega}{\omega_{0}}}
        {\hat a}\left({\bf
        k_{0}},\bar{q}_{-}\right){\hat b}\right\},
        \label{eq:Heff0}
\end{eqnarray}
where now $\bar{q}_{\pm}=\sqrt{(\omega_{0} \pm \Omega)^{2}/c^{2}-k_{0}^{2}}$.
The physical process described by this interaction Hamiltonian is
very similar to a stimulated Brillouin scattering \cite{PER84}, even though in
this case the Stokes and anti-Stokes component are back-scattered by
the acoustic wave at reflection, and the optomechanical coupling is provided by the
radiation pressure and not by the dielectric properties of the mirror (see Fig.~1 for a schematic
description).

\begin{figure}
\begin{center}
\includegraphics[width=0.26\textwidth]{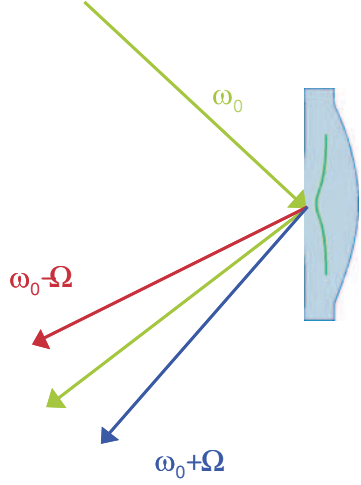}
\end{center}
\vspace{-0.5cm} \caption{\label{setup} Schematic description of
the system. A laser field at frequency $\omega_{0}$ impinges on
the mirror vibrating at frequency $\Omega$. In the reflected field
two sideband modes are excited at frequencies
$\omega_{1}=\omega_{0}-\Omega$ (Stokes mode) and
$\omega_{2}=\omega_{0}+\Omega$ (antiStokes mode). }
\end{figure}

In practice, either the driving laser beam and the back-scattered modes
are never monochromatic, but have a nonzero bandwidth. In general the
bandwidth of the back-scattered modes is determined by the bandwidth
of the driving laser beam and that of the acoustic mode. However, due
to its high mechanical quality factor, the spectral width of the
mechanical resonance is negligible (about $1$ Hz) and, in practice, the
bandwidth of the two sideband modes $\Delta \nu_{mode}$
coincides with that of the incident laser beam.
It is then convenient to consider this nonzero bandwidth to redefine
the bosonic operators of the Stokes and anti-Stokes modes
to make them dimensionless,
\begin{eqnarray}
{\hat a}_{1}&=& 2\pi \sqrt{\frac{2\pi \Delta \nu_{mode}}{cA}}
{\hat a}\left({\bf k_{0}},\bar{q}_{-}\right) =
2 \pi \sqrt{\frac{\Delta q}{A}}{\hat a}\left({\bf k_{0}},\bar{q}_{-}\right)\\
{\hat a}_{2}&=& 2\pi \sqrt{\frac{2\pi \Delta \nu_{mode}}{cA}}
{\hat a}\left({\bf k_{0}},\bar{q}_{+}\right)=
2 \pi \sqrt{\frac{\Delta q}{A}}{\hat a}\left({\bf k_{0}},\bar{q}_{+}\right),
\end{eqnarray}
so that  Eq.(\ref{eq:Heff0}) reduces to an effective
Hamiltonian
\begin{equation}
    {\hat H}_{eff}=-i\hbar \chi
    ({\hat a}_{1}{\hat b}-{\hat a}^{\dag}_{1}{\hat b}^{\dag})
    -i\hbar\theta({\hat a}_{2}{\hat b}^{\dag}-{\hat a}^{\dag}_{2}
    {\hat b})\,,
    \label{eq:Heff}
\end{equation}
where the couplings $\chi$ and $\theta$ are given by
\begin{eqnarray}
\label{chi}
\chi &=& q_{0}\Delta\nu_{det}\sqrt{\frac{\hbar}{2M\Omega}}
\sqrt{\frac{{\wp} }{ \Delta \nu_{mode}\hbar \omega_{0}}}
\sqrt{\frac{\omega_{0}-\Omega}{\omega_{0}}}=
\cos\phi_{0}\sqrt{\frac{{\wp}\Delta\nu_{det}^{2}(\omega_{0}-\Omega)}{2M\Omega
c^{2}\Delta\nu_{mode}}} \\
\theta &=& \chi \sqrt{\frac{\omega_{0}+\Omega}{\omega_{0}-\Omega}},
\end{eqnarray}
with $\phi_{0}=\arccos(cq_{0}/\omega_{0})$, is the angle of
incidence of the driving beam. It is possible to verify that with
the above definitions, the Stokes and anti-Stokes annihilation
operators $\hat{a}_{1}$ and $\hat{a}_{2}$ satisfy the usual
commutation relations
$\left[\hat{a}_{i},\hat{a}_{j}^{\dagger}\right]=\delta_{i,j}$.

\section{System dynamics}

Eq.~(\ref{eq:Heff}) shows that the radiation pressure of an intense monochromatic laser beam incident on a mirror
generates an effective coupling between a given vibrational mode
of the mirror and the two first optical sideband modes induced by the vibrations. One has a system of three, bilinearly coupled, bosonic modes, which is formally
equivalent to a system of three optical modes interacting through coupled parametric down-conversion and up-conversion with simultaneous phase
matching. This system has been experimentally studied in the classical domain in \cite{rabin}, while the quantum photon statistics has been
determined in Ref.~\cite{smithers}. More recently, the same optical system has been proposed for the implementation of symmetrical
and asymmetrical quantum telecloning in \cite{paris1}. The same three mode dynamics has also been studied within a cavity configuration \cite{wong}, where
however the dissipation associated with cavity losses significantly changes the dynamics. Finally, a similar Hamiltonian describes also
the interaction of two atomic modes with opposite momentum in a Bose-Einstein condensate with an off-resonant optical
mode scattered by it, and its quantum dynamics has been recently discussed in \cite{paris2}.

Eq.~(\ref{eq:Heff}) contains two interaction terms: the first one,
between modes ${\hat a}_{1}$ and ${\hat b}$,
is a parametric-type interaction
leading to squeezing in phase space \cite{QO94}, and it is
able to generate the EPR-like
entangled state which has been used in the CV teleportation
experiment of Ref.~\cite{FUR98}. The
second interaction term, between modes ${\hat a}_{2}$ and ${\hat b}$,
is a beam-splitter-type
interaction \cite{QO94}. This purely Hamiltonian description
satisfactorily reproduces the dynamics as long as the dissipative
coupling of the mirror vibrational mode with its environment is negligible.
This happens for vibrational modes with a high-Q mechanical quality factor.
In this case we can consider an interaction time, i.e., a
time duration of the incident laser pulse, much smaller than the
relaxation time of the vibrational mode (which can be of order of one
second \cite{TIT99}).

The Hamiltonian (\ref{eq:Heff}) leads to a system of
linear Heisenberg equations, namely
\begin{eqnarray}
    \dot{\hat a}_{1}&=&\chi {\hat b}^{\dag}\,,
    \label{eq:sol1} \\
    \dot{\hat b}&=&\chi {\hat a}_{1}^{\dag}-\theta {\hat a}_{2}\,,
    \label{eq:sol2} \\
    \dot {\hat a}_{2}&=&\theta {\hat b}
\label{eq:sol3} \,.
\end{eqnarray}
The solutions read
\begin{eqnarray}
    {\hat a}_{1}(t)&=&\frac{1}{\Theta^{2}}\left[\theta^{2}-\chi^{2}
    \cos\left(\Theta t\right)\right]{\hat a}_{1}(0)
    +\frac{\chi}{\Theta}\sin\left(\Theta t\right) {\hat b}^{\dag}(0)
    -\frac{1}{\Theta^{2}}\left[\chi\theta-\chi\theta
    \cos\left(\Theta t\right)\right]{\hat a}_{2}^{\dag}(0)\,,
    \\
    {\hat b}(t)&=&-\frac{\chi}{\Theta}\sin\left(\Theta t\right)
    {\hat a}_{1}^{\dag}(0)
    +\cos\left(\Theta t\right) {\hat b}(0)
    -\frac{\theta}{\Theta}\sin\left(\Theta t\right) {\hat a}_{2}(0)\,,
    \\
    {\hat a}_{2}(t)&=&\frac{1}{\Theta^{2}}\left[\chi\theta-\chi\theta
    \cos\left(\Theta t\right)\right]{\hat a}_{1}^{\dag}(0)
    -\frac{\theta}{\Theta}\sin\left(\Theta t\right) {\hat b}(0)
    -\frac{1}{\Theta^{2}}\left[\chi^{2}-\theta^{2}
    \cos\left(\Theta t\right)\right]{\hat a}_{2}(0)\,,
\end{eqnarray}
where $\Theta=\sqrt{\theta^{2}-\chi^{2}}$.

On the other hand, the system dynamics can be easily studied also through
the normally ordered characteristic function $\Phi(\mu,\nu,\zeta)$ \cite{QO94},
where $\mu,\nu,\zeta$ are the complex variables corresponding
to the operators ${\hat a}_{1},{\hat b},{\hat a}_{2}$ respectively.
From the Hamiltonian (\ref{eq:Heff}) the dynamical equation for
$\Phi$ results
\begin{eqnarray}\label{eq:Phidot}
    {\dot\Phi}&=&\chi\left(
    \mu\nu+\mu^{*}\nu^{*}-\mu^{*}\frac{\partial}{\partial\nu}
    -\mu\frac{\partial}{\partial\nu^{*}}
    -\nu^{*}\frac{\partial}{\partial\mu}
    -\nu\frac{\partial}{\partial\mu^{*}}\right)\Phi
    \nonumber\\
    &&+\theta\left(
    \zeta^{*}\frac{\partial}{\partial\nu^{*}}
    +\zeta\frac{\partial}{\partial\nu}
    -\nu^{*}\frac{\partial}{\partial\zeta^{*}}
    -\nu\frac{\partial}{\partial\zeta}\right)\Phi\,,
\end{eqnarray}
with the initial condition
\begin{equation}\label{eq:Phiini}
    \Phi(t=0)=\exp\left[-\overline{n}|\nu|^{2}\right]\,,
\end{equation}
corresponding to the vacuum for the modes ${\hat a}_{1}$,
${\hat a}_{2}$ and to a thermal state for the mode ${\hat b}$.
The latter is characterized by an average number of excitations
$\overline{n}=[\coth(\hbar\Omega/2k_{B}T)-1]/2$,
$T$ being the equilibrium temperature and $k_{B}$ the
Boltzmann constant.
Since the initial condition is a Gaussian state of the three-mode system,
and the system is linear, the joint state of the whole system at time
$t$ is still Gaussian, with characteristic function
\begin{equation}\label{eq:Phisol}
    \Phi=\exp\left[
    -{\cal A}|\mu|^{2}-{\cal B}|\nu|^{2}-{\cal E}|\zeta|^{2}
    +{\cal C}\mu\nu+{\cal C}\mu^{*}\nu^{*}
    +{\cal F}\mu\zeta+{\cal F}\mu^{*}\zeta^{*}
    +{\cal D}\nu\zeta^{*}+{\cal D}\nu^{*}\zeta\right]\,,
\end{equation}
where
\begin{eqnarray}
    {\cal A}(t)&=&\frac{1}{2(r^2-1)^2}\left\{
    \left[1-\cos\left(2\Theta t\right)\right]\left[\overline{n}(r^2-1)-1\right]
    \right.\nonumber\\
    &&\left.
    +4r^2 \left[1-\cos\left(\Theta t\right)\right]\right\}\,,
    \label{eq:A}
    \\
    {\cal B}(t)&=&\frac{1}{2(r^2-1)}\left\{
    \left[1+\cos\left(2\Theta t\right)\right]\overline{n}(r^2-1)
    +1-\cos\left(2\Theta t\right)\right\}\,,
    \label{eq:B}
    \\
    {\cal C}(t)&=&\frac{1}{2(r^2-1)^{3/2}}\left\{
    2r^2
    \sin\left(\Theta t\right)\right.
    \nonumber\\
    &&\left.
    +\left[\overline{n}(r^2-1)-1\right]
    \sin\left(2\Theta t\right)\right\}\,,
    \label{eq:C}
    \\
    {\cal D}(t)&=&\frac{-r}{2(r^2-1)^{3/2}}\left\{
    2\sin\left(\Theta t\right)
    +\left[\overline{n}(r^2-1)-1\right]
    \sin\left(2\Theta t\right)\right\}\,,
    \label{eq:D}
    \\
    {\cal E}(t)&=&\frac{r^2}{2(r^2-1)^2}\left\{
    \left[1-\cos\left(2\Theta t\right)\right]
    \left[\overline{n}(r^2-1)-1\right]\right.
    \nonumber\\
    &&\left.
    +4\left[1-\cos\left(\Theta t\right)\right]\right\}\,,
    \label{eq:E}
    \\
    {\cal F}(t)&=&\frac{r}{2(r^2-1)^2}\left\{
    \left[1-\cos\left(2\Theta t\right)\right]
    \left[\overline{n}(r^2-1)-1\right]\right.
    \nonumber\\
    &&\left.
    +2(1+r^2) \left[1-\cos\left(\Theta t\right)\right]\right\}\,,
    \label{eq:F}
\end{eqnarray}
and $r=\theta/\chi=\left[(\omega_{0}+\Omega)/(\omega_{0}-\Omega)\right]^{1/2}$. Eqs.~(\ref{eq:A}-\ref{eq:F}) show that the system dynamics
depends upon three dimensionless parameters:
the initial vibrational mean thermal excitation number $\bar{n}$; the scaled time $t'\equiv\Theta t$; the ratio between coupling constants $r$, which depends only upon the ratio between the
vibrational frequency and the optical driving frequency and it is always very close to one. Since within the approximations
introduced in the preceding Section, the system can be described as a closed system of three interacting oscillators, the resulting dynamics is periodic, with period
$2\pi$ for the scaled variable $t'$. For this reason we shall consider only $0\leq t'\leq 2\pi $ in the following.

The state of the system can be equivalently described also in terms of its correlation matrix (CM) $V$, defined by $V_{ij}=\langle \Delta
\hat{\xi}_{i}\Delta\hat{\xi}_{j}+\Delta\hat{\xi}_{j}\Delta
\hat{\xi}_{i}\rangle /2$, where $\Delta\hat{\xi}_{i}=\hat{\xi}_{i}-\langle \hat{\xi}_{i}\rangle
$ and $\hat{\xi}$ denotes the following vector of quadratures, $\hat{\xi}=(\hat{X}_{1},\hat{P}_{1},\hat{X}_{b}, \hat{P}_{b},\hat{X}_{2},\hat{P}_{2})$,
defined by $\hat{X}_{k} =(\hat{a}_{k}+\hat{a}_{k}^{\dagger})/\sqrt{2}$, $\hat{P}_{k}=(\hat{a}_{k}-\hat{a}_{k}^{\dagger})/i\sqrt{2}$ ($k=1,2$), and
$\hat{X}_{b}=(\hat{b}+\hat{b}^{\dagger})/\sqrt{2}$, $\hat{P}_{b}=(\hat{b}-\hat{b}^{\dagger})/i\sqrt{2}$. In fact, the CM of a Gaussian state
is directly connected with the
symmetrically ordered correlation function $\Phi_{sym}(\mu,\nu,\zeta) = \exp\left\{-(|\mu|^{2}+|\nu|^{2}+|\zeta|^{2})/2\right\} \Phi(\mu,\nu,\zeta)$.
In the present case, the three mode system is at all times a zero-mean Gaussian state and therefore this connection can be written as
\begin{equation}
\Phi_{sym}(\xi)= \exp\left\{-\sum_{i,j=1}^{6}\xi_{i}V_{ij}\xi_{j}\right\},
\label{symme}
\end{equation}
where $\xi$ is the six-dimensional vector of phase space coordinates corresponding to the quadratures $\hat{\xi}$. Using Eq.~(\ref{eq:Phisol}), one gets
\begin{equation}
V=\left(
\begin{tabular}
[c]{cc|cc|cc}
${\cal A}+\frac{1}{2}$ & $0$ & ${\cal C}$ & $0$ & ${\cal F}$ & $0$\\
$0$ & ${\cal A}+\frac{1}{2}$ & $0$ & $-{\cal C}$ & $0$ & $-{\cal F}$\\\hline
${\cal C}$ & $0$ & ${\cal B}+\frac{1}{2}$ & $0$ & $-{\cal D}$ & $0$\\
$0$ & $-{\cal C}$ & $0$ & ${\cal B}+\frac{1}{2}$ & $0$ & $-{\cal D}$\\\hline
${\cal F}$ & $0$ & $-{\cal D}$ & $0$ & ${\cal E}+\frac{1}{2}$ & $0$\\
$0$ & $-{\cal F}$ & $0$ & $-{\cal D}$ & $0$ & ${\cal E}+\frac{1}{2}$\\
\end{tabular}
\right) \;.
\label{CMgrande}
\end{equation}

\section{Separability properties of the tripartite CV system}

In the preceding Section we have completely determined the dynamics of the system and these results allow us to perform a complete
characterization of the separability properties of this tripartite CV system. For systems composed of more than two
parties, one has many ``types'' of entanglement due to the many ways in which the different subsystems may be entangled with each other.
As a first task we classify the entanglement possessed by the system as a function of the various parameters.

\subsection{Entanglement classification}

Following the scheme introduced in \cite{dur}, one has {\em five} different entanglement classes for a tripartite system:

{\em Class 1}. Fully inseparable states, i.e. not separable for any grouping of the parties.

{\em Class 2}. One-mode biseparable states, which are separable if two of the parties are grouped together, but inseparable with respect to the other groupings.

{\em Class 3}. Two-mode biseparable states, which are separable with respect to two of the three possible bipartite splits
but inseparable with respect to the third.

{\em Class 4}. Three-mode biseparable states, which are separable with respect to all three bipartite splits but cannot be written as a
mixture of tripartite product states.

{\em Class 5}. Fully separable states, which can be written as a mixture of tripartite product states.

The state of our tripartite system is easy to determine at some specific time instants. Due to periodicity,
at $t'= 0, 2 \pi$ the state is simply the chosen initial state, i.e.
the tripartite product of a thermal state for the mirror and the vacuum state for the Stokes and the antiStokes sideband modes. Therefore at
these isolated time instants the state belongs to class $5$ simply because of the assumed fully separable initial state.
The state of the system is easy to determine also at
$t'= \pi$, i.e. just in the middle of its periodic evolution.
In fact, choosing $t'= \pi$ in Eq.~(\ref{CMgrande}), the CM becomes
\begin{equation}
V\left(t'=\pi\right)  =\left(
\begin{tabular}
[c]{cc|cc|cc}%
$\frac{4r^{2}}{\left(  r^{2}-1\right)  ^{2}}+\frac{1}{2}$ & $0$ & $0$ & $0$ &
$\frac{2r\left(  r^{2}+1\right)  }{\left(  r^{2}-1\right)  ^{2}}$ & $0$\\
$0$ & $\frac{4r^{2}}{\left(  r^{2}-1\right)  ^{2}}+\frac{1}{2}$ & $0$ & $0$ &
$0$ & $-\frac{2r\left(  r^{2}+1\right)  }{\left(  r^{2}-1\right)  ^{2}}%
$\\\hline
$0$ & $0$ & $\bar{n}+\frac{1}{2}$ & $0$ & $0$ & $0$\\
$0$ & $0$ & $0$ & $\bar{n}+\frac{1}{2}$ & $0$ & $0$\\\hline
$\frac{2r\left(  r^{2}+1\right)  }{\left(  r^{2}-1\right)  ^{2}}$ & $0$ & $0$%
& $0$ & $\frac{4r^{2}}{\left(  r^{2}-1\right)  ^{2}}+\frac{1}{2}$ & $0$\\
$0$ & $-\frac{2r\left(  r^{2}+1\right)  }{\left(  r^{2}-1\right)  ^{2}}$ & $0$%
& $0$ & $0$ & $\frac{4r^{2}}{\left(  r^{2}-1\right)  ^{2}}+\frac{1}{2}$%
\end{tabular}
\right)= V_{12}^{TMS}\oplus V_{b}^{\bar{n}},
\label{pi}
\end{equation}
where $V_{12}^{TMS}$ is the $4 \times 4$ CM of a TMS state
for the two optical sidebands and $V_{b}^{\bar{n}}$ is the one-mode CM of a thermal state with mean excitation number $\bar{n}$ for the mirror.
Therefore at half period the mirror vibrational
mode factorizes from the two optical modes which are instead in a TMS state with a two-mode squeezing
parameter given by $\sinh^{-1}\left[2r(r^{2}+1)/(r^{2}-1)^{2}\right]$. This means that at $t'=\pi$ the state is a one-mode biseparable
state, i.e. belongs to class $2$, for any value of the mirror temperature and of the ratio $r$.

It is important to notice that the reduced state of the two
sideband modes at $t'=\pi$ is independent of the mirror temperature and, being a TMS state, its entanglement can be
arbitrarily increased by increasing the squeezing parameter,
which in this case is realized by letting $r \to 1 \Leftrightarrow \Omega/\omega_0 \to 0$.
This means that when the interaction time, i.e. the duration of the driving laser pulse, is carefully chosen so that $t'=\pi$, the
two reflected sideband modes are entangled in the same way as the twin beams at the output of a nondegenerate parametric amplifier
and this entanglement is insensitive with respect to the mirror temperature. Therefore, under appropriate conditions, radiation pressure becomes
an alternative tool for the generation of CV entanglement between optical modes \cite{JOPB}.

\begin{figure}[htbp]
\begin{center}
\includegraphics[width=3.0in,height=2.16in]{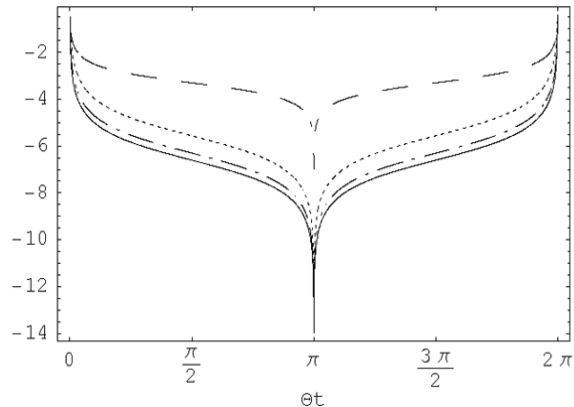}
\end{center}
\caption{Logarithmic negativity $\log_{10}\left\{|\eta_{b}(t')|\right\}$ (see text) vs the scaled time $t'=\Theta t$.
The values of other parameters are: $r=1+2.5\times10^{-7}$ and $\bar{n}=0$ dashed line, $\bar{n}=10^{-1}$ dotted line, $\bar{n}=1$ dotted-dashed line,
$\bar{n}=10^{7}$ solid line.}
\label{fig01}
\end{figure}

For interaction times different from the above special cases, we can still determine the entanglement class of the system state
by applying the results of Ref.~\cite{entcirac}, which has provided a necessary and sufficient criterion for the determination of the class in the case
of tripartite CV Gaussian states and which is directly computable. This classification criterion is mostly based on the
nonpositive partial transposition (NPT) criterion proved in \cite{entwerner}, which is necessary and sufficient for $ 1\times N$ bipartite CV Gaussian
states. The NPT criterion of \cite{entwerner} can be expressed in terms of the symplectic matrix
\begin{equation}
{\cal J} = \bigoplus_{i=1}^3 J_i, \;\;\;\;\;\;
J_i=\left(
\begin{array}
[c]{cc}
0 & 1\\
-1 & 0
\end{array}
\right) \;\;\;\;\; {\rm i=1,2,3} \label{sympl}
\end{equation}
expressing the commutation rules between the canonical coordinates $
[\hat{\xi}_{\alpha},\hat{\xi}_{\beta}]  =i{\cal J}_{\alpha\beta}$, and of the partial transposition transformation $\Lambda_k$, acting
on system $k$ only.
Transposition is equivalent to time reversal and therefore in phase space is equivalent to change the sign of the momentum operators, i.e.
$\Lambda_k (\ldots, X_k, P_k, \ldots) = (\ldots, X_k, -P_k, \ldots)$. The NPT criterion states that
a $1\times N$ CV Gaussian state is separable if and only if the ``test matrix'' $\tilde{V}_{k}=\Lambda_{k}V\Lambda
_{k}+i{\cal J}/2 \geq 0$. Therefore by evaluating the sign of the eigenvalues of the three matrices $\tilde{V}_{k}$, $k=1,2,b$ and using the
NPT criterion one can discriminate states belonging to class $1$ (none of the three matrices $\tilde{V}_{k}$ is positive), class $2$ (only
one of the $\tilde{V}_{k}$ is positive), and class $3$ (two of the $\tilde{V}_{k}$ are positive). When all the $\tilde{V}_{k}$ are positive, NPT criterion
is not able to distinguish between class $4$ and $5$ and a specific criterion for this case has been derived in \cite{entcirac}. However in the present case
this latter result is not needed and the entanglement classification is simple also for interaction times $t' \neq m \pi$ (integer $m$). In fact,
denoting with $\eta_{j}(t',r,\bar{n})$ the minimum eigenvalue of $\tilde{V}_{j}$, we have numerically checked that $\eta_{j}(t',r,\bar{n})$
is always negative for $j=1,2,b$, $\forall t' \neq m \pi$ and for a very large range of $\bar{n}$ and $r$ \cite{footn},
and therefore the state of the system is fully inseparable.
For example, in the case $r=1+2.5\times10^{-7}$, $\forall t' \neq m \pi$ we
have $\eta_{1}(t',\bar{n})  \simeq \eta_{2}(t',\bar{n})\simeq -0.5$ independent of $\bar{n}$, with only $\eta_{b}(t',\bar{n})$
significantly depending on the mirror temperature. By taking the logarithm of the modulus of this negative eigenvalues as a measure
of entanglement (a sort of logarithmic negativity \cite{loga}), this means that only the entanglement between the vibrational mode and the subsystem
formed by the two optical modes is sensitive to temperature, while the entanglement between one sideband mode and the other two modes is not.
This example of logarithmic negativity $\log_{10}\left\{|\eta_{b}(t')|\right\}$
is shown in Fig.~\ref{fig01} for different values of $\bar{n}$. As expected, the entanglement between the mirror mode and the two optical mode decreases with
increasing temperature.
We conclude that our tripartite state is almost everywhere fully
entangled (class 1) except for the isolated
time instants $t'= 2 k \pi $ when it is fully separable (class 5) and
at instants $t'=(2k+1) \pi $ when it is one-mode biseparable (class 2).

\subsection{Entanglement of the bipartite reduced systems}

It is also interesting to study the entanglement properties of the bipartite system which is obtained
when one of the three modes is traced out. This study may be interesting for possible applications of the
present simple optomechanical scheme for quantum information processing.

The Gaussian nature of the state is preserved after the partial trace operation and the CM of the bipartite reduced state
is immediately obtained from the
original three-mode CM of Eq.~(\ref{CMgrande}) by eliding the blocks relative to the traced out mode.
To study the entanglement of the three Gaussian bipartite states we
use the Simon's necessary and sufficient criterion for entanglement \cite{entsimon}, which is just the NPT criterion
discussed above in the special case of $1\times 1$ CV Gaussian states. This criterion can be expressed in a compact form
valid for all the three possible bipartite reduced systems by redefining the index $j=b$ as $j=3$ and defining
$$
\begin{array}
[c]{cccccc}%
\alpha_{1}=\alpha_{4}\equiv {\cal E} & \alpha_{2}\equiv {\cal B} & \alpha_{3}\equiv {\cal A} &
\alpha_{5}\equiv {\cal D} & \alpha_{6}\equiv {\cal C} & \alpha_{7}\equiv {\cal F}.
\end{array}
$$
One has three markers of entanglement $\Upsilon^{(j)}$ ($j$ denotes the traced mode) and the state of the
corresponding bipartite system is entangled when
\begin{eqnarray}
\Upsilon^{(j)}\equiv\left(  \alpha_{j}\alpha_{j+1}+\frac{1}{4}+\frac
{\alpha_{j}+\alpha_{j+1}}{2}-\alpha_{j+4}^{2}\right)  ^{2}+\frac{1}{16}-\frac{\alpha_{j+4}^{2}}{2}\nonumber\\
-\left(  \frac{\alpha_{j}}{2}+\frac{1}{4}\right)  ^{2}-\left(  \frac
{\alpha_{j+1}}{2}+\frac{1}{4}\right)  ^{2}<0.
\label{criterio}
\end{eqnarray}

First of all one has $\Upsilon^{(1)}(t')\geq 0$ $\forall t',\bar{n}$, i.e.
the antiStokes mode $\hat{a}_{2}$\ and the mirror vibrational mode are never entangled.
This is not surprising if we take into account that the two modes interact with each other with the beam-splitter-like
interaction proportional to $\chi$ in Eq.~(\ref{eq:Heff}) and that, as shown in \cite{kim},
nonclassical input states are needed for a beam splitter to generate entanglement between two modes.
Since the chosen initial state is classical, we do not expect to have any entanglement between the antiStokes sideband and the vibrational mode.

\begin{figure}[htbp]
\begin{center}
\includegraphics[width=3.0in,height=2.16in]{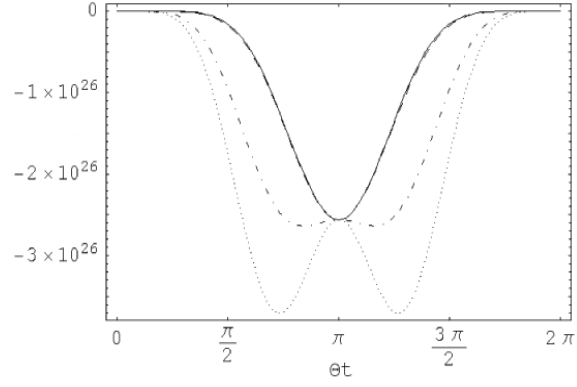}
\end{center}
\caption{Marker of entanglement $\Upsilon^{(3)}$ between the two sideband modes vs scaled time $t^{\prime}\equiv\Theta t$. The two modes are
always entangled except for $t'=0,2 \pi$.
The values of other parameters are: $r=1+2.5\times10^{-7}$ and $\bar{n}=0$ (solid line), $\bar{n}=10^{5}$ (dashed line), $\bar{n}=5\times10^{6}$ (dotted-dashed line),
$\bar{n}=10^{7}$ (dotted line).}
\label{fig02}
\end{figure}

On the contrary, if we instead trace out the vibrational mode, we find $\Upsilon^{(3)}(t')  <0$ , $\forall
t'\neq 0,2\pi$ and for a wide range of mirror temperatures, from $\bar{n}=0$ to $n = 10^{7}$.
This means that the entanglement between the two optical modes, created by
the ponderomotive interaction with the vibrating mirror, is extremely robust with respect to the
mirror temperature (see Fig.~\ref{fig02}). This is an indirect effect of the one-mode biseparable state assumed by the system at half period
$t'=\pi$ discussed in the preceding subsection, in which the two sideband modes are in a TMS state unaffected by the mirror
temperature. Departing from the $t'=\pi$ condition, the reduced state of the reflected modes becomes temperature-dependent and is no more
a pure TMS state. However the two modes remain entangled and this entanglement persists even at very high temperatures. Therefore
an interesting result of our analysis is that the radiation pressure of an intense driving beam incident on a vibrating mirror
is able to entangle in a robust way the two reflected sideband modes \cite{JOPB}.

\begin{figure}[htbp]
\begin{center}
\includegraphics[width=3.0in,height=2.16in]{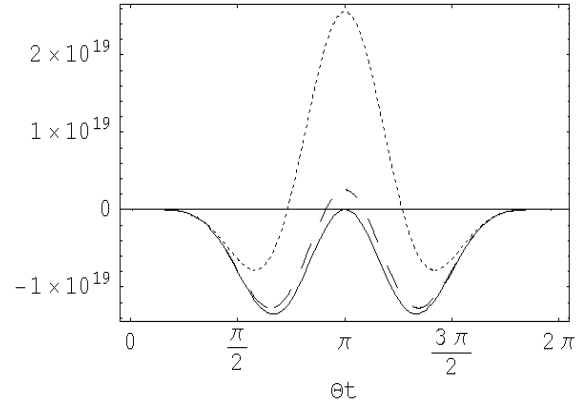}
\end{center}
\caption{Marker of entanglement $\Upsilon^{(2)}$ between the Stokes sideband and the vibrational mode
vs scaled time $t^{\prime}\equiv\Theta t$. The values of other parameters are: $r=1+2.5\times10^{-7}$ and $\bar{n}=0$ (solid line), $\bar {n}=10^{-8}$ (dashed line),
$\bar{n}=10^{-7}$ (dotted line).}
\label{fig03}
\end{figure}

\begin{figure}[htbp]
\begin{center}
\includegraphics[width=3.0in,height=2.16in]{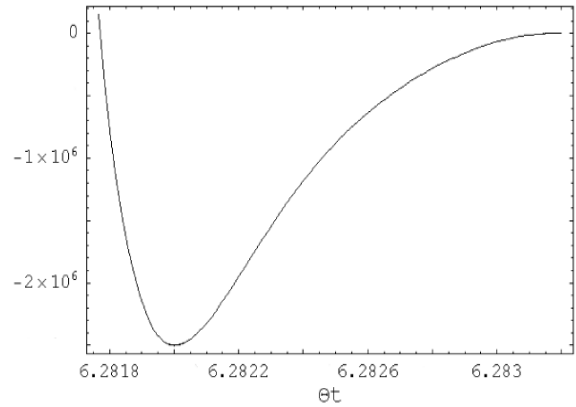}
\end{center}
\caption{Marker of  entanglement $\Upsilon^{(2)}$ between the Stokes sideband and the vibrational mode
vs scaled time $t^{\prime}\equiv\Theta t$ around $t' = 2 \pi$. The values of other parameters are: $r=1+2.5\times10^{-7}$ and $\bar{n}=10^{3}$.
This shows that entanglement between the two modes persists even at high temperatures in this small time interval. }
\label{fig04}
\end{figure}

Finally, if we trace out the antiStokes mode $a_2$ we still find time intervals in which the Stokes mode and the vibrational mode
are entangled, i.e. $\Upsilon^{(2)}(t')  <0 $ , but now this entanglement is very sensitive to the mirror temperatures. In fact.
the time intervals for which $\Upsilon^{(2)}(t')  <0 $ becomes soon very narrow
even for very small but nonzero $\bar{n}$ (see Fig.~\ref{fig03}). This is consistent with the sensitivity upon temperature of the logarithmic negativity shown
in Fig.~\ref{fig01}. It is important to notice however that, even for
large values of $\bar{n}$, a small time interval around $t' = 2\pi $ of negative values of $\Upsilon^{(2)}(t')$
still exists (see Fig.~\ref{fig04}).
This small interval of interaction times where the mirror vibrational mode and the Stokes sideband mode remain entangled
even at high temperatures is very important
because such an entanglement is necessary for the implementation of any scheme able to transfer CV quantum information
between optical modes and the vibrational mode of the mirror. In the next section we shall see how this optomechanical entanglement
can be used to teleport an unknown CV quantum state of an optical mode onto the vibrational mode of a mirror.

\section{Teleportation onto the state of the vibrational mode}

\begin{figure}[ht]
\begin{center}
\includegraphics[width=3.0in]{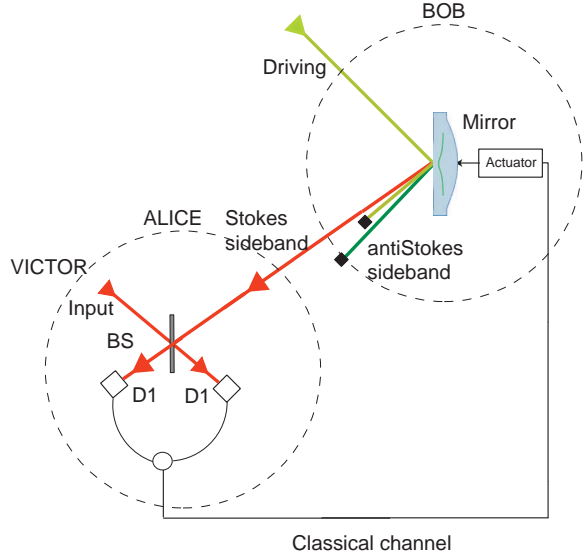}
\end{center}
\caption{Scheme for the teleportation of an unknown quantum state
of the optical mode provided by Victor to Alice onto the
vibrational mode of the mirror. Only the Stokes sideband reaches
Alice's station and is mixed in the 50-50 beam splitter BS with
the unknown input given by Victor. A Bell-like measurement $D1$
using homodyne detection is then performed on this combination and
the result is fed-forward to Bob through the classical channel.
Finally, he actuates the phase space displacement on the moving
mirror. } \label{fig1}
\end{figure}

We have seen that radiation pressure realizes an effective coupling between optical and acoustic modes. Since optical travelling beams are
the typical playground for CV quantum information applications, this fact gives the possibility to involve also the vibrational modes
of a mirror in the manipulation and storage of CV quantum information. The first experimental demonstration of a CV quantum information protocol has been the
quantum teleportation of an unknown coherent state of an optical mode onto another optical mode illustrated in \cite{FUR98} and recently
refined in \cite{BOW03}. We shall discuss how this CV quantum teleportation protocol can be adapted to the optomechanical system
studied here in order to realize the teleportation of an unknown state of an optical field onto a vibrational mode of the mirror.
This will allow us to present and discuss in more detail the optomechanical teleportation scheme of \cite{PRL03}.
Quantum teleportation requires the use of shared entanglement between two distant stations, Alice and Bob, and of a classical channel for the transmission
of the results of the Bell measurement from Alice to Bob \cite{BEN93}.
The entanglement required is therefore that between the mirror vibrational mode and one optical mode. However, it is evident from the Hamiltonian
of Eq.~(\ref{eq:Heff}) and the bipartite entanglement analysis of the preceding Section that the relevant optical sideband mode is the Stokes mode.
In fact, it interacts with the vibrational mode through the same parametric interaction generating two-mode squeezing which is at the basis of
the optical version of the CV quantum teleportation protocol proposed in \cite{BRA98} and realized in \cite {FUR98,BOW03}. With this respect, the presence
of the additional beam-splitter-like interaction with the antiStokes mode can only degrade the bipartite entanglement between the Stokes mode and
the vibrational mode.
The most straightforward way to implement teleportation in this optomechanical setting is just to apply the protocol of
\cite{BRA98} and ignoring (i.e. tracing out) the disturbing antiStokes sideband mode. A schematic description of this solution
is shown in Fig.~\ref{fig1}. An unknown
quantum state of a radiation field is prepared by a verifier
(Victor) and sent to Alice, where it is mixed at a balanced beam splitter with the Stokes sideband mode with frequency $\omega_0-\Omega$.
This latter mode is the only one reaching Alice station after reflection upon the mirror because the driving beam and the antiStokes mode are filtered out.
The radiation pressure of the driving beam on the mirror establishes the required quantum channel, i.e. the shared entangled state
between the mirror vibrational mode in Bob's hand and the Stokes sideband at Alice station. It is evident from the analysis of Sections III and IV
that this bipartite reduced state is still Gaussian, with CM
\begin{equation}
    V_{1b}=\left(
    \begin{array}{cccc}
    {\cal A}+\frac{1}{2}&0
    &{\cal C}&0
    \\
    0&{\cal A}+\frac{1}{2}
    &0&-{\cal C}
    \\
    {\cal C}&0
    &{\cal B}+\frac{1}{2}&0
    \\
    0&-{\cal C}
    &0&{\cal B}+\frac{1}{2}
    \end{array}
    \right)\,.
    \label{eq:Gam1}
\end{equation}

Then Alice carries out a balanced homodyne detection at each output port of the beam splitter, thereby
measuring two commuting quadratures
${\hat X}_{+}=({\hat X}_{in}+{\hat X}_{1})/\sqrt{2}$ and
${\hat P}_{-}=({\hat P}_{in}-{\hat P}_{1})/\sqrt{2}$, with
measurement results $ X_{+}$ and $P_{-}$.
The final step at the sending station is to transmit the classical
information, corresponding to the result of
the homodyne measurements she
performed, to the receiving terminal.
Upon receiving this information, Bob displaces his part of entangled
state (the mirror acoustic mode) as follows:
${\hat X}_{b}\to {\hat X}_{b}+\sqrt{2}X_{+}$,
${\hat P}_{b}\to {\hat P}_{b}-\sqrt{2}P_{-}$.
To actuate the phase-space displacement,
Bob can use again the radiation pressure force.
In fact, if the mirror is shined by a bichromatic intense laser
field with frequencies $\varpi_{0}$ and
$\varpi_{0}+\Omega$, it is easy to understand that, after averaging over the rapid oscillations,
the radiation pressure force yields an effective interaction Hamiltonian
$H_{act} \propto {\hat b}e^{-i\varphi}+{\hat b}^{\dag}e^{i\varphi}$,
where $\varphi$ is the relative phase between the two frequency
components. Any phase space displacement of the mirror vibrational
mode can be realized by adjusting this relative phase and the
intensity of the laser beam.

\begin{figure}[ht]
\begin{center}
\includegraphics[width=3.2in]{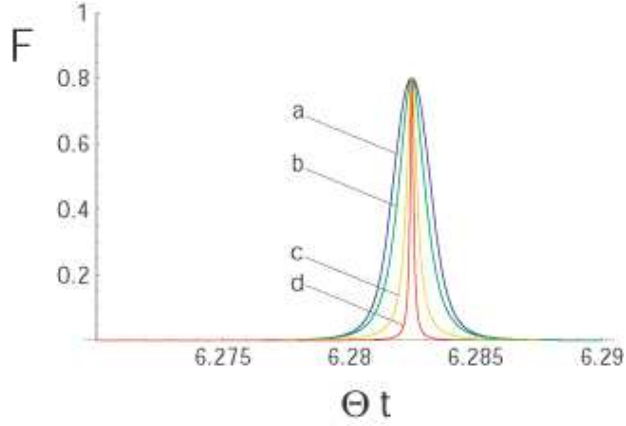}
\end{center}
\caption{Fidelity $F$ vs the scaled time $t'=\Theta t$ for the
teleportation protocol of Fig.~\protect\ref{fig1} in which the
antiStokes mode is traced out. Curves a, b, c, d are for
$\overline{n}=0$, 1, 10, $10^{3}$, respectively.} \label{fig2}
\end{figure}

In the case when Victor provides an input Gaussian state
characterized by a $2\times 2$ CM $V^{in}$,
the output state at Bob site is again Gaussian, with
CM $V^{out}$.
The input-output relation for these matrices can be found as
follows. In terms of symmetrically ordered characteristic functions we have
\begin{equation}
    \exp\left[-\sum_{i,j=1}^2 u_{i}V^{out}_{ij}u_{j}\right]
    =\tilde{K}(u_1, u_2)
    \exp\left[-\sum_{i,j=1}^2u_{i}V^{in}_{ij}u_{j}\right]
\end{equation}
where ${\bf u}$ is the variable vector of the characteristic
functions and $\tilde{K}(u_1,u_2)$ can be written in terms of the symmetrically ordered characteristic function
of the entangled state shared by Alice and Bob $\Phi_{sym}^{1b}(\mu,\nu)$ as
\begin{eqnarray}
    \tilde{K}(u_1, u_2)&=&
    \int d^2k d^2\lambda d^2\mu d^2\nu \Phi_{sym}^{1b}(\mu,\nu) \exp\left\{-ik_1 u_1-ik_2 u_2+i\mu_1 \lambda_1 -i\mu_2 \lambda_2+i\nu_1(k_1-\lambda_1)+i\nu_2(k_2-\lambda_2)
    \right\}
    \nonumber\\
    &=&\exp\left[-\sum_{i,j=1}^4 w_i
    V^{1b}_{ij}w_j\right]\,,
\end{eqnarray}
where ${\bf w} = (u_1,-u_2,u_1,u_2)$. Then, it is easy to derive the input-output relation for these matrices (see also \cite{CHI02})
\begin{eqnarray}
    V_{11}^{out}&=&V_{11}^{in}+\left(
    V^{1b}_{11}+2V^{1b}_{13}+V^{1b}_{33}\right)\,,
    \label{eq:Gout1} \\
    V_{12}^{out}&=&V_{12}^{in}+\left(
    V^{1b}_{14}-V^{1b}_{12}+V^{1b}_{34}-V^{1b}_{23}\right)\,,
    \\
    V_{22}^{out}&=&V_{22}^{in}+\left(
    V^{1b}_{22}-2V^{1b}_{24}+V^{1b}_{44}\right)\,. \label{eq:Gout3}
\end{eqnarray}
The fidelity $F$ of the described teleportation protocol,
which is the probability to obtain the input state at Bob site, can be written, with the
help of Eqs.~(\ref{eq:Gout1})-(\ref{eq:Gout3}) and (\ref{eq:Gam1}), as
\begin{equation}
    F=\{1+\left[1+{\cal A}+{\cal B}+2{\cal C}\right]\}^{-1}
    \label{eq:Fid}
\end{equation}
where we have specialized to the case of an input coherent state.
In such a case, the upper bound for the fidelity achievable with only
classical means and no quantum resources
is $F=1/2$ \cite{BRA99}. Fig.~\ref{fig2}
shows the fidelity as a function of the interaction time $t'$ only in the short time interval around $t'= \Theta t = 2\pi$
where it is appreciably different from zero.
Different values of the initial mean thermal phonon number of
the mirror acoustic mode $\overline{n}$ are considered ($\overline{n}=0$,
1, 10, $10^{3}$), and we have fixed again $r=1+2.5\times10^{-7}$,
as we have done in the preceding Section. The fidelity is appreciably different from zero only for interaction times $t'$ slightly smaller than
$2\pi$ and this time region corresponds just to the interaction time interval considered in Fig.~\ref{fig04} where the entanglement between the Stokes mode and the
vibrational mode persists even at high mirror temperatures.
The interesting result shown in Fig.~\ref{fig2}
is that $F$ reaches a maximum value, $F_{max} \simeq 0.80$, which is well above the classical bound $F=0.5$ and that it
is surprisingly independent of the initial temperature of the acoustic mode.
This effect could be ascribed to quantum interference
phenomena, and opens the way
for the demonstration of quantum teleportation of states
of mesoscopic massive systems composed of very many atoms.
However the mirror temperature still has important disturbing effects,
since by increasing $\overline{n}$,
the useful time interval over which $F>1/2$ soon becomes extremely narrow, similarly to what happens to the
visibility in the interference experiment proposed in \cite{marshall} to detect the linear superposition of two macroscopically distinct
states of a mirror.
In the present proposal that means the need of designing precise driving laser pulses
in order to have a well defined interaction time corresponding to the maximum of fidelity. This is however feasible using
available pulse shaping techniques, since what is required here is simply a control of the laser pulse duration at the level
of hundreds of nanoseconds.

\begin{figure}[ht]
\begin{center}
\includegraphics[width=3.0in]{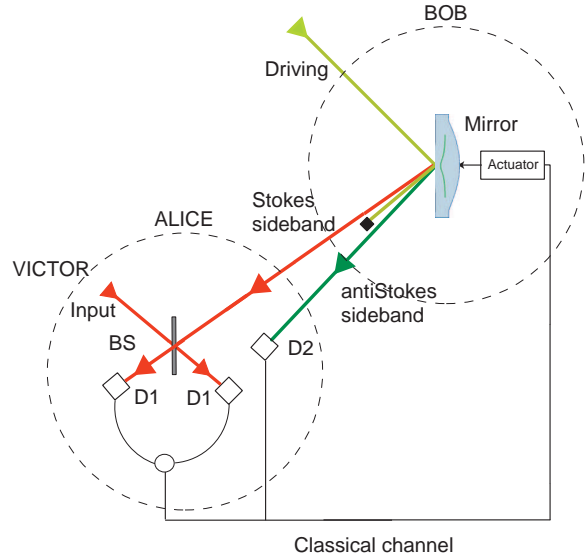}
\end{center}
\caption{Improved scheme for the teleportation of an unknown
quantum state of the optical mode provided by Victor to Alice onto
the vibrational mode of the mirror. {\em Both} the Stokes and the
antiStokes modes reach Alice's station. The Stokes mode is mixed
in the 50-50 beam splitter BS with the unknown input given by
Victor as in the previous scheme of Fig.~\protect\ref{fig1}. The
antiStokes mode is instead subject to a heterodyne measurement.
The results of both the homodyne and heterodyne measurements are
sent to Bob through the classical channel and he then uses this
information to perform the appropriate phase-space displacement of
the mirror. } \label{fighet}
\end{figure}

The results of Fig.~\ref{fig2} refer however to the straightforward procedure of ignoring the antiStokes mode. The fact that
the teleportation process is not optimal, i.e. $F$ never reaches $1$, even asymptotically, can be ascribed to the fact that
this ``spectator'' sideband mode carries away part of the quantum correlations which could be useful for teleportation.
One could try therefore to adapt the original teleportation protocol of \cite{BRA98} to the present situation. A modified
CV quantum teleportation protocol has been introduced in \cite{PRL03}: the antiStokes mode is not traced out but it is sent to Alice site
as the Stokes mode and it is then subjected to a heterodyne measurement \cite{YUE80} (see Fig.~(\ref{fighet})). Such a measurement serves the
purpose of acquiring as much quantum
information as possible about the tripartite entanglement between the two sideband modes and the vibrational mode and using it in a
profitable way for teleportation. Performing an additional heterodyne measurement at Alice site certainly improves the protocol
with respect to the scheme discussed above discarding the second sideband. Moreover it has the additional advantages of being easily
implementable experimentally and of preserving the Gaussian nature of the entangled state of the Stokes and vibrational modes.
On the other hand we have not proved that this is the best measurement Alice can perform on the $a_2$ mode to improve teleportation
and therefore one cannot exclude that Alice, with appropriate measurements and local operations could get an even better result.

When Alice performs the heterodyne measurement
on the mode ${\hat a}_{2}$, the latter is projected onto a coherent
state of complex amplitude $\alpha$ \cite{YUE80}.
The Gaussian entangled state
for the optical Stokes mode ${\hat a}_{1}$ and the vibrational mode $\hat{b}$,
is now conditioned to this measurement result, which however explicitly appears only in first order moments, i.e. only
the mean values of the quadratures. The CM of the resulting entangled state is instead independent of $\alpha$,
and it is given by
\begin{equation}
    V_{1b}^{het} =
    \left(
    \begin{array}{cccc}
    {\cal A}+\frac{1}{2} -\frac{{\cal F}^{2}}{{\cal E}+1}&0
    &{\cal C}+\frac{\cal FD}{{\cal E}+1}&0
    \\
    0&{\cal A}+\frac{1}{2}-\frac{{\cal F}^{2}}{{\cal E}+1}
    &0&-{\cal C}-\frac{\cal FD}{{\cal E}+1}
    \\
    {\cal C}+\frac{\cal FD}{{\cal E}+1}&0
    &{\cal B}+\frac{1}{2}-\frac{{\cal D}^{2}}{{\cal E}+1}&0
    \\
    0&-{\cal C}-\frac{\cal FD}{{\cal E}+1}&
    0&{\cal B}+\frac{1}{2}-\frac{{\cal D}^{2}}{{\cal E}+1}
    \end{array}
    \right)\,.
    \label{eq:Gam2}
\end{equation}
Then the protocol proceeds in the same way as above. Alice carries out the homodyne measurement of the commuting quadratures
${\hat X}_{+}$ and ${\hat P}_{-}$, with
measurement results $ X_{+}$ and $P_{-}$ and then she sends the information
corresponding to the results of {\em both} homodyne and heterodyne measurements to Bob, who now displaces his part of entangled
state as follows:
${\hat X}_{b}\to {\hat X}_{b}+\sqrt{2}X_{+}
+\sqrt{2}{\rm Re}\{\alpha\}({\cal F}-{\cal D})/({\cal E}+1)$,
${\hat P}_{b}\to {\hat P}_{b}-\sqrt{2}P_{-}
+\sqrt{2}{\rm Im}\{\alpha\}({\cal F}+{\cal D})/({\cal E}+1)$.
Notice that now Bob's local operation depends
on all Alice's measurement results ($X_{+}$, $P_{-}$, $\alpha$).
The final phase-space displacement of the mirror is realized in the same way as before.
Also the input-output relation between the CM of the input state provided by Victor and the output state in Bob's hands
is the same as above, except that now the $4\times 4$ CM matrix of the shared entangled state
$V^{1b}$ of Eq.~(\ref{eq:Gam1}) is replaced by the CM $V^{het}_{1b}$ of Eq.~(\ref{eq:Gam2}) in Eqs.~(\ref{eq:Gout1})-(\ref{eq:Gout3}).

\begin{figure}[ht]
\begin{center}
\includegraphics[width=3.0in]{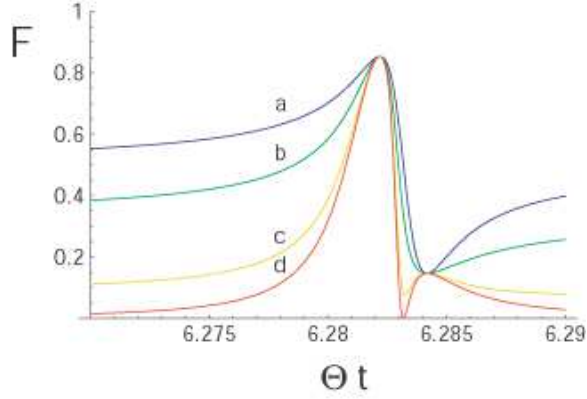}
\end{center}
\caption{Fidelity $F$ vs the scaled time $t'\equiv\Theta t$ for
the improved teleportation protocol of Fig.~\protect\ref{fighet}
with the additional heterodyne measurement of the antiStokes mode.
Curves a, b, c, d are for $\overline{n}=0$, 1, 10, $10^{3}$,
respectively.} \label{fig3}
\end{figure}

The fidelity $F$ of this modified protocol, referred again to the case of an input coherent state, can be written as
\begin{equation}
    F=\{1+\left[1+{\cal A}+{\cal B}+2{\cal C}
    -({\cal F}-{\cal D})^{2}/({\cal E}+1\right)]\}^{-1}\;,
    \label{eq:Fid2}
\end{equation}
and its behavior in the same interaction time interval around $t'= 2\pi$
and for the same four different values of $\bar{n}$ considered in Fig.~\ref{fig2} is shown in Fig.~\ref{fig3}.
Comparing Figs.~\ref{fig2} and \ref{fig3}, one sees that
the improvement brought by this second protocol exploiting the additional heterodyne measurement is evident.
Due to quantum interference, one has again a temperature-independent maximum fidelity, which is however
larger than before, since it is $F_{max} \simeq 0.85$. The modified protocol is much more robust with respect to temperature
and the time interval over which the fidelity is appreciably nonzero is now significantly larger, even though, at high temperatures,
the time interval in which one has quantum teleportation, i.e. $F> 1/2$, is still quite narrow and an accurate shaping of the driving
laser pulse is still needed.

\begin{figure}[ht]
\begin{center}
\includegraphics[width=3.0in]{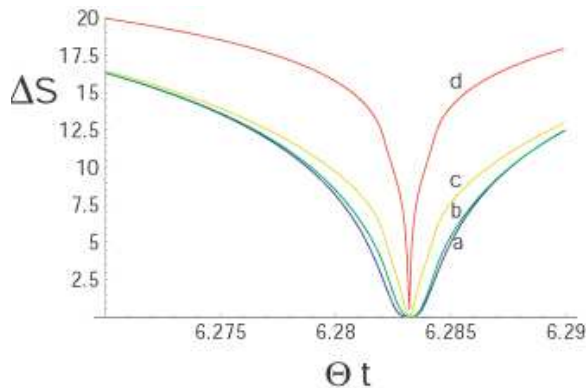}
\end{center}
\caption{Information gain $\Delta S$ brought by the heterodyne
measurement vs the scaled time $t^{\prime}\equiv\Theta t$ in the
same small time interval around $t=2 \pi$ considered in
Figs.~\protect\ref{fig2} and \protect\ref{fig3} for different
values of the mirror mean vibrational number $\bar{n}$. Curves a,
b, c, d are for $\overline{n}=0$, 1, 10, $10^{3}$, respectively.}
\label{fig4}
\end{figure}

A quantitative comparison of the improvement brought by the additional heterodyne measurement on the antiStokes mode to the
teleportation protocol can be made by evaluating the additional quantum information acquired by Alice with this measurement.
This information can be evaluated in terms of the von Neumann entropies of the involved bipartite entangled states shared by Alice and Bob.
In fact, we have seen that the effect of the heterodyne measurement on the antiStokes mode is to change the CM of the entangled state
used as quantum channel for the teleportation. Therefore the quantum information gained by Alice with the heterodyne measurement is
given by
\begin{equation}
\Delta S = S_{1b}-S_{1b}^{het}
\label{info}
\end{equation}
where $S_{1b}= -$Tr$\left\{\rho_{1b}\log\rho_{1b}\right\}$ is the von Neumann entropy of the Gaussian entangled state of the Stokes mode
and the vibrational mode when the antiStokes mode is traced out, with CM given by Eq.~(\ref{eq:Gam1}) and
$S_{1b}^{het}= -$Tr$\left\{\rho_{1b}^{het}\log\rho_{1b}^{het}\right\}$ is the von Neumann entropy of the bipartite state conditioned to
the result of the heterodyne measurement and with CM given by Eq.~(\ref{eq:Gam2}). These two entropies can be easily evaluated in terms of the
CM of the two bipartite entangled states. In fact, using the results of \cite{serafini} one has that the von Neumann entropy $S(V)$ of a bipartite
Gaussian state with $4\times 4$ CM $V$, which is divided in $2\times 2$ blocks $A$, $B$ and $C$ as
\begin{equation}
    V=\left(
    \begin{array}{cc}
    A& C
    \\
    C^T& B
    \\
    \end{array}
    \right)\,,
    \label{eq:VS}
\end{equation}
can be written as
\begin{equation}
S(V)=\sum_{i=+,-}\left[\left(n_i+\frac{1}{2}\right)\log\left(n_i+\frac{1}{2}\right)-\left(n_i-\frac{1}{2}\right)\log\left(n_i-\frac{1}{2}\right)\right]\;,
\label{entro}
\end{equation}
where
\begin{equation}
n_{\pm}=\sqrt{\frac{\Delta(V)\pm \sqrt{\Delta(V)^2-4 {\rm Det}V}}{2}}
\label{ennei}
\end{equation}
and
\begin{equation}
\Delta(V)={\rm Det} A +{\rm Det} B+2 {\rm Det} C\;.
\end{equation}
The behavior of the quantum information gained by Alice $\Delta S$ of Eq.~(\ref{info}) versus the interaction time $t'$, in the same small time interval
around $t'=2\pi$ considered in Figs.~\ref{fig2} and \ref{fig3} and for different values of the initial mean vibrational number $\bar{n}$
is plotted in Fig.~\ref{fig4}. We see that the information acquired through the heterodyne measurement increases for larger temperatures, i.e.,
the modified protocol becomes particularly useful at higher temperatures. This is in agreement with the fact that the modified protocol
with the additional heterodyne measurement is much more robust with respect to temperature than the one based on tracing out the antiStokes mode, as showed by
Figs.~\ref{fig2} and \ref{fig3}. In Fig.~\ref{fig4} we see also that the quantum information gain $\Delta S$ drops to zero at all temperatures
at exactly $t'=2 \pi$, because at this interaction time the bipartite reduced state of the Stokes and the vibrational modes becomes the chosen factorized initial state
of Eq.~(\ref{eq:Phiini}), i.e. the product of a
thermal equilibrium state for the vibrational mode and the vacuum state for the optical sideband, either with or without the additional heterodyne measurement.

\section{Discussion of the results and concluding remarks}

We have seen that the optomechanical entanglement between the Stokes sideband mode and a vibrational mode of a mirror can be profitably used
for the teleportation of an arbitrary state of a radiation mode onto the quantum state of the vibrational mode.
This would imply the remarkable capability of manipulating CV quantum information with objects with a macroscopic number of atoms.
For example, the possibility to ``write'' the state of an optical beam onto an acoustic mode implies the possibility to perform
a very fast ground state cooling of the mirror vibrational mode. In fact, in this case, cooling is nothing but the teleportation
of the optical vacuum state provided by Victor onto the state of the mirror vibrational mode. This ``telecooling'' is very fast
and it exploits the optomechanical entanglement created by radiation pressure.
As a matter of fact, the effective number of thermal excitations of the vibrational mode soon after
the two homodyne and the heterodyne measurements at Alice station (in the improved protocol of Fig.~\ref{fighet}) becomes
$\overline{n}_{eff}=1+{\cal A}+{\cal B}+2{\cal C}
-({\cal F}-{\cal D})^{2}/({\cal E}+1)$.
It reduces to $\overline{n}+1$ in absence of entanglement,
where $1$ represents the noise introduced by the protocol.
Instead, the optomechanical interaction for a proper time permits
to achieve $\overline{n}_{eff}=0.17$
at once, at the moment of Alice's measurement, thanks to the nonlocal quantum correlations between the mirror mode and the light
mode at Alice's station.

A remarkable aspect of the present scheme is the robustness of the tripartite entanglement between the two motional sidebands
and the vibrational mode with respect to the mirror temperature. In fact, the three modes are almost always entangled, at
any temperature, and this holds provided that the interaction time, i.e. the duration of the driving laser pulse
is much smaller than the mirror mode decay rate. In this case, the coupling of the mirror with its environment can be neglected
and the dynamics of the three-mode system is purely Hamiltonian, so that
quantum coherence effects can still survive even for a highly mixed mirror initial state.
On the other hand, thermal noise has still important effects: i) the interaction time window
in which the protocol is efficient becomes narrower and narrower for increasing temperature (see the previous Section);
ii) the time interval within
which the classical communication from Alice to Bob, and the phase
space displacement by Bob have to be made, becomes shorter and shorter
with increasing temperature, because the vibrational state projected
by Alice's Bell measurement heats up in a time of the order
of $(\gamma_{m}\overline{n})^{-1}$,
where $\gamma_{m}$ is the mechanical damping constant.
Therefore, despite this robustness, in practice, any experimental implementation will still need an
acoustic mode cooled at low temperatures (see however
Refs.~\cite{COH99,VIT02}
for effective cooling mechanism of acoustic modes).

The effects of mechanical damping can be neglected during the
back-scattering process stimulated by the intense laser beam when the parameter $\Theta$, whose inverse fixes the dynamical timescale of
the process, is much larger than the decay rate $\gamma_m$.
Mechanical damping rates of about $\gamma_{m} \simeq 1$ Hz are
available \cite{TIT99}, and therefore choosing
$\chi \simeq \theta \simeq 5\times 10^{5}$ Hz, and $r -1 = 2.5 \times 10^{-7}$ one has
$\Theta=\sqrt{\theta^{2}-\chi^{2}}\simeq
10^{3}$ Hz, which means using a highly monochromatic laser pulse with a duration of milliseconds.
Such values for the coupling constants $\chi$ and $\theta$ can be obtained for example choosing
${\wp} = 10$W, $\omega_{0}\sim 2\times 10^{15}$ Hz, $\Omega \sim
5\times 10^{8}$ Hz, $\Delta\nu_{det}\sim 10^{7}$ Hz,
$\Delta\nu_{mode}\sim t^{-1} \sim 10^{3}$ Hz, and $M \sim 10^{-10}$
Kg. These parameters
are partly different from those of already
performed optomechanical experiments \cite{TIT99,COH99}.
However, using a thinner silica crystal and considering higher
frequency modes, the parameters we choose could be obtained. On the other hand
performing a quantum teleportation experiment on a collective mode of a mirror involving about $10^{15}$ atoms
is not trivial at all and the technical difficulties in the realization of this experiment are not surprising.
Such low effective masses of the involved vibrational mode could be achieved for example
in micro-opto-electro-mechanical systems (MOEMS), which possess vibrational modes
with a very high oscillation frequency $\Omega$ \cite{ROU03}.

There are also other sources of imperfection
besides the thermal noise acting on
the mechanical resonator. Other fundamental noise sources are
the shot noise and the radiation pressure noise.
Shot noise affects the detection of the sideband modes but its effect
is already taken into account by our treatment (it is responsible for the $1/2$ terms in the
diagonal entries of the correlation matrix of Eqs.~(\ref{eq:Gam1}) and (\ref{eq:Gam2})).
The radiation pressure noise due to the intensity fluctuations of the
incident laser beam yields fluctuations of the optomechanical
coupling. However, the chosen parameter values
give $\Delta\wp/\wp = \Delta (\Theta t)/2\Theta t \simeq
10^{-8}$, which is negligibly small with respect to the width of the
window where the fidelity attains its maximum (see Figs.~\ref{fig2} and \ref{fig3}).

Finally, one has also to verify that the
teleportation has been effectively realized, that is, that the state provided by Victor has been
transferred to the mirror vibrational mode. This implies measuring the final state of the
acoustic mode, and this can be done
by considering a second, intense ``reading'' laser pulse,
and exploit again the optomechanical interaction given by
Eq.~(\ref{eq:Heff}), where now $a_{1}$ and $a_{2}$ are output meter modes.
It is in fact possible to perform a heterodyne measurement
of an
appropriate combination of the two back-scattered modes,
${\hat Z}={\hat a}_{1}-{\hat a}_{2}^{\dag}$, if
the driving laser beam at frequency $\omega_0$ is used as local
oscillator and the resulting photocurrent is mixed with a signal oscillating
at the frequency $\Omega$.
The behaviour of $Z(t)$ as a function of the time duration of the
second ``measuring'' driving beam can be derived from Eqs.~(\ref{eq:sol1})-(\ref{eq:sol3}),
that is
\begin{eqnarray}
{\hat Z}(t)\equiv {\hat a}_{1}(t)-{\hat a}_{2}^{\dag}(t)&=&
\frac{1}{\Theta}\left[\chi+\theta\right]\sin(\Theta t)
{\hat b}^{\dag}(0)
\nonumber\\
&+&\frac{1}{\Theta^{2}}\left[
\theta^{2}-\chi^{2}\cos(\Theta t)-\chi\theta+\chi\theta\cos(\Theta
t)\right]{\hat a}_{1}(0)
\nonumber\\
&-&\frac{1}{\Theta^{2}}\left[
\chi\theta+\chi\theta\cos(\Theta t)-\chi^{2}-\theta^{2}\cos(\Theta
t)\right]{\hat a}_{2}^{\dag}(0)\,.
\end{eqnarray}
It is easy to see that for $\cos(\Theta t)=0$ and
$\Theta(\theta+\chi)\gg\theta(\theta-\chi)$
the measured quantity practically coincides with the mode
oscillation operator $b^{\dag}(0)$, thus revealing information on the
state of the mechanical oscillator.

To summarize, we have seen that the radiation pressure of an intense and quasi-monochromatic laser pulse
incident on a perfectly reflecting mirror
is able to entangle a mirror vibrational mode with the two first optical sidebands of the driving mode.
The three-mode CV system shows almost always fully tripartite entanglement for any value of the mirror temperature,
provided that the coupling of the mirror mode with its environment can be neglected during the
optomechanical interaction. In particular, the bipartite entanglement between the Stokes sideband mode and the mirror vibrational mode
can be used to implement the teleportation of an unknown quantum state of a radiation field
onto a macroscopic, collective vibrational
degree of freedom of a massive mirror.

The present result could be challenging tested
with present technology, and opens new perspectives towards the use
of quantum mechanics in macroscopic world.


\begin{thebibliography}{99}
\bibitem{footn}Actually we have only considered
values of $r$ very close to $1$, because they are the only ones which are physically significant.
\end{thebibliography}
\end{document}